# A block-based inter-band predictor using multilayer propagation neural network for hyperspectral image compression

Rui Dusselaar and Manoranjan Paul, *Senior Member, IEEE*

*Abstract*— In this paper, a block-based inter-band predictor (BIP) with multilayer propagation neural network model (MLPNN) is presented by a completely new framework. This predictor can combine with diversity entropy coding methods. Hyperspectral (HS) images are composed by a series high similarity spectral bands. Our assumption is to use trained MLPNN predict the succeeding bands based on current band information. The purpose is to explore whether BIP-MLPNN can provide better image predictive results with high efficiency. The algorithm also changed from the traditional compression methods encoding images pixel by pixel, the compression process only encodes the weights and the biases vectors of BIP-MLPNN which require few bits to transfer. The decoder will reconstruct a band by using the same structure of the network at the encoder side. The BIP-MLPNN decoder does not need to be trained as the weights and biases have already been transmitted. We can easily reconstruct the succeeding bands by using the BIP-MLPNN decoder. The experimental results indicate that BIP-MLPNN predictor outperforms the CCSDS-123 HS image coding standard. Due to a good approximation of the target band, the proposed method outperforms the CCSDS-123 by more than 2.0dB PSNR image quality in the predicted bands. Moreover, the proposed method provides high quality image e.g., 30 to 40dB PSNR at very low bit rate (less than 0.1 bpppb) and outperforms the existing methods e.g., JPEG, 3DSPECK, 3DSPIHT and in terms of rate-distortion performance.

## INTRODUCTION

A TYPICAL visible light spectrum for human-vision is about 400 nm to 700 nm in wavelength. Hyperspectral (HS) images contain an extensive range of spectral information which can provide a rich observation power beyond the capability of human vision. HS image can be considered as the integration of digital imaging with spectroscopy. It has the ability to capture more information than a standard (RGB) images. The contained spectral detail in HS images gives better capability to characterize the objects. HS spectral information also can be used to analyse moisture content, texture, reflectance and other external quality characteristics. The extensive implementations have been used in both the civilian and military field, such as satellite/airborne based remote sensing [1], target detection [2, 3], non-invasive quality inspection [4], classification [5] as well as quality control in food and agriculture [6, 7] and other lab applications etc.

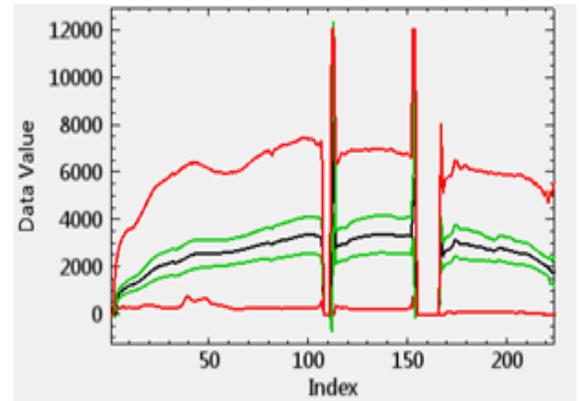

Fig. 1. Four-pixel values are shown against 224 different spectral bands with wavelengths from 0.4 to 2.5 μm of the Cuprite 2 image of NASA's AVIRIS dataset to demonstrate the wide range of values and variations in different bands which make HS image compression challenging. The red curves at bottom and top are minimum and maximum reflectance values respectively, the curve in black is mean of green pixels vectors.

In Fig. 1, an example of an HS image from NASA's AVIRIS dataset is shown. This HS image carries 224 contiguous spectral bands with wavelengths from 0.4-2.5 μm. It shows that each pixel $P = (p_1, p_2,... p_n)$ is represented as a vector with 224 elements. Because of carrying wealthy spectral information, HS image data become huge with very high redundancy. Those pixel vectors have properties of nonlinearity, continuity and similarity. Especially similarity trend in neighbouring bands of the spectrum can be extracted as learning samples to train a neural network. This point has been well proven in the experiments results of this paper.

The unique structure of HS images has resulted in its compression method different from the commonly used image compression methods or video encoding methods because of no

R. Dusselaar is with the Data Science Research Centre, School of computing and Mathematics, Charles Sturt University, Bathurst, NSW 2795, Australia (e-mail: rxiao@csu.edu.au). M. Paul is with the Data Science Research Centre, School of Computing and Mathematics, Charles Sturt University, Bathurst, NSW 2795, Australia (e-mail: mpaul@csu.edu.au).

motion. Both spectral and special redundancy need to be fully considered.

The predictive encoding is one of the basic techniques realizing image compression. Using intra-band, inter-band, or hybrid predictors to decorrelate the redundancy of HS images is a very active research direction. Regardless of which type of predictors have been chosen, an ideal intra/inter-band predictor is the maximum approximation to the actual target data then followed by entropy coding. The encoding and decoding process only act on residuals, consequently reduces the computational complexity and transmitted cost.

Currently, available literature presents various methods regarding HS image compression studies. The method to remove band-to-band redundant information can achieve a significant compression ratio [8]. Band selection-based (BS) method is to select a subset of bands from HS image. Zhao *et al*. [9] introduce an algorithm based on intra-band prediction and inter-band fractal encoding. HS bands are partitioned into several groups of bands (GOBs). The authors apply intra-band prediction to the first band in each GOB. The hypothesis is that two blocks (8x8 pixels) located in the same position of adjacent HS bands are highly similar. Likewise, the authors of [10] observe that HS images have strong similarities between adjacent bands. They use the first band for intra-band prediction encoded and the remaining bands are inter-band encoded using fractal encoded. The encoding process of fractal coding is to use the similar interpolation method. A more efficient GOBs selection is vital for the algorithm performance. There is no universal metric of GOBs that is applicable to all HS images in different wavelengths. Moreover, in some experimental results [9, 10] show that the algorithms can only achieve effective compression at a low bit rate. The methods in [11, 12] use a two-stage predictor: one is an inter-band linear predictor, and the other based on least square predictor. The two-stage predictor can remove redundancy from two directions, however, the computation cost is also relatively higher in this technique.

Prediction-based techniques often use a mathematical model to predict pixel values and encode only their prediction residuals [26]. DPCM is an important prediction approach. DPCM includes a linear predictor and median filter. The predictor calculates the residual between the value of the current pixel and the predicted pixel. The residual normally has a smaller variance. It results in fewer bits for coding the image. An improved DPCM [13], named as C-DPCM, uses separated spectral clusters. The mean-square error inside each cluster is used to calculate the coefficients. All the pixels used to make the prediction have the same spatial location as the current pixel, and then the difference between the actual targeted value and the predicted value is encoded. This algorithm provides easy mathematical derivation and computational advantage. However, in many practical applications this still shows an undesirable predictive result.

More commonly, compression techniques divided into two main categories: lossless and lossy compression methods, depending on whether the original image can be precisely re-generated from the compression data [14]. Lossless compression is used for applications that require the reconstructed image to restore to the original signal with high precision. Because of the intrinsic entropy of the data, lossless compression algorithm generally achieves modest compression ratio and cost more storage space. By contrast, a lossy compression algorithm is for preserving essential spectral information of target objects, which gives a balance of compression efficiency and loss of information. Lossy image compression mainly uses predictive coding methods to reduce redundancy among bands and transform coding methods to compact important information for compression. Lossy compression is used when the user can tolerate some signal loss. The lossless compression methods are generally included dictionary-based schemes and statistical schemes [15].

Statistical-based schemes require distribution knowledge where the compression takes place based on the frequency of input characters. The methods in [16, 17] encoded HS images based on LookUp Tables (LUT). The LUT searches the previous band for a pixel equal to the current band in the same position called a predictor. The predictor is used as a key to search LUT to speed up the search process. The most well-known statistical-based algorithms are Huffman Coding [18] and Arithmetic Coding [19, 20].

Transform-based technique, such as the Pairwise Orthogonal Transform (POT), also called multiple pairwise PCA [21] is one of the spectral transforms that an image is transformed using multiple pairwise operations instead of a single transform. It overcomes the problem of KLT, such as bit depth expansion, lack of scalability and reduced memory requirements [22]

Recently, a method in [23] proposed a Gaussian mixture-based modelling technique to predict the succeeding band from current bands. The predicted band is then used as the additional reference band along with the previous band to apply on high efficiency video coding standard (HEVC). Using the number of Gaussian distributions and the initial parameters setting is vital for the result accuracy.

Exploiting the special data structure of an HS image, a number of 3D transform-based compression methods have been proposed including 3D-SPECK which has been applied to an HS image to exploit the joint properties of the spatial and spectral correlations [24]. 3D-SPIHT is named as the benchmark for 3D image compression. The tree structure of 3DSPIHT Zala et al. [25] and Zayed et al. [26] extended 2D-SPIHT tree structure to a third dimension. Tang et al. [24] proved that that 3D-SPECK is better than 3D-SPIHT to achieve an efficient compression. AT-3DSPECK (asymmetric transform 3DSPECK) was introduced by Tang et al. [27] which is a more efficient tree structure. The tree structure is with longer depth and therefore to provide better energy concentration. Wu et al. [28] extended the *Context-Based Adaptive Lossless Image Codec* (CALIC) algorithm from 2D to 3D-CALIC. Besides, the algorithm in [29] used 3D wavelet for denoising to achieve good performance. 3-D context-based adaptive lossless image coding (M-CALIC) Magli et al. [30] made another extension of CALIC, it modified inter-band predictor and used thresholds in quantization. It's also a commonly used benchmark encoder and yields better coding performance than CCSDS-123.

Tensors had been broadly used in physics and engineering applications. Recently some researchers have branched it out to the image processing area. A tensor can be understood as arrays in multidimensional space. It was introduced by Hitchcock in 1927 [31]. Latterly the famous Tucker model was discovered [32]. It can be considered the same as PCA on high-dimensional data and keep the spatial structure of the data. Tucker decomposition decomposes a tensor into a set of matrices and one small core tensor. Karami *et al*. [33] newly applied a nonnegative Tucker decomposition. HS image is treated as a 3-D tensor and spatially partitioned into smaller sub-tensors. Another most widely used tensor decomposition is PARAFAC [34] or CANDECOMP [35] that decomposes a tensor as a sum of rank-one tensors. Those two types of tensors models are higher-order extensions of the matrix singular value decomposition, they can also be considered as a higher-order form of PCA. Hang *et al*. [36] treated HS image as a 3-order-tensor. Original data is decomposed into a core tensor. By this way, it can accomplish the purpose of lower dimensionality. This algorithm gets a slightly higher false alarm rate for target detection performance and complexity in the calculation. A solution has been proposed in [37]. Veganzones *et al*. employed a compression-based nonnegative canonical polyadic decomposition algorithm to reduce memory requirements and to speed up computations. [38] propose a compression method using patch-based low-rank tensor decomposition. The third-order tensor was from local patch. the similar tensor patches were grouped to form a fourth-order tensor.[39, 40] used Low-rank Matrix for HS image denoising.It is worth mentioning that the Consultative Committee for Space Data Systems (CCSDS) have published new compression standard for HS data s. The core predictor in CCSDS called "Fast Lossless (FL)" developed by the NASA Jet Propulsion Laboratory.

By studying on inter-band prediction encoding, if the prediction accuracy is high enough, the amount of transmitted data could be reduced dramatically. The artificial neural network based coding technique for image compression research is very active and fast development. Its predictive ability can be made use for predictive coding. Especially BIP-MLPNN with back propagation training algorithm is one of the most popular neural network algorithms and has been used largely in image processing [41, 42]. The method in [43] predicts the band-wise correlation of HS images based on a generalized regression neural network.

The main novelty of the proposed approach is to utilize the excellent nonlinear approximation capability of artificial neural network (ANN) to develop a high accuracy predictor, preferably approximate the relationships between the current band and the succeeding band for the purpose of inter-band prediction. The compression process only encodes the weights and the biases vectors of neural network which require few bits to transfer. Thus, we achieve significant image quality improvement compared to the existing methods.

## Methodology

The multilayer feed-forward network can be used to approximate almost arbitrary curves if we have enough neurons in the hidden layers [44]. The neural network stores the specific information in the weights and biases of the network. The current spectral band is used as a training data set, and the next band is the training target set in the encoder side. Then weights and biases need to be encoded by the binary entropy encoder. In the decoder side, the next (target) band can be reconstructed by using transferred weights, biases, residual based on the immediately previous band information. This is equivalent to representing the original sample with a smaller data which is actually a compression process.

Fig. 2 illustrates the prediction process of the proposed algorithm. It consists of six main parts: data normalization; BIP-MLPNN encoder; entropy; BIP-MLPNN decoder; images reconstruction and reverse normalization. The step-by-step operations are described below:

**Step 1**.  Pre-process, resize and rescale image data to a 256×256 matrix and a range between [0, 1]

**Step 2**.  BIP-MLPNN modelling. Use a transfer function *tansig* as hidden layer and the linear function *purelin* as the output layer.

**Step 3.**  Tune the values of the weights and biases of the network to approximate the expected value.

**Step 4.**  Encode the weights, biases and compensation residuals if needed. Send to decoder.

**Step 5.**  Bands reconstruction by using BIP-MLPNN decoder. MLPNN decoder is the same structure network as encoder side. MLPNN decoder can reconstruct the image by using function *tansig* and *purelin* with the weights, biases and the previous band as input data.

**Step 6.**  Reverse normalization, mapping minimum and maximum values to [0, 255] data space.

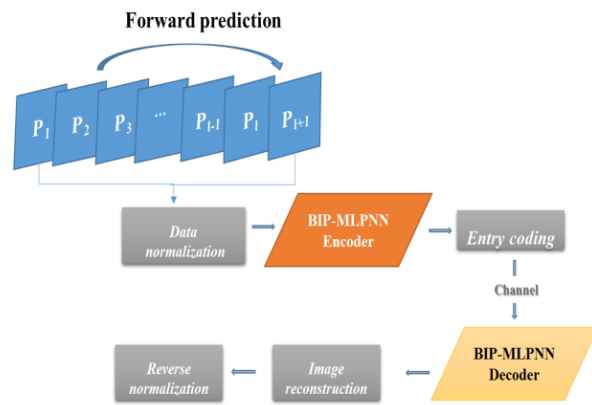

Fig. 2.  A simplified prediction process by the proposed algorithm. it illustrates the prediction process

The first nonzero band is encoded as lossless binary codes then transmitted to the decoder. The BIP-MLPNN will be *train*ed on *the enc*oder side and then send weights and biases to the decoder side. Each upcoming band needs to be trained once. The prediction process is an iterative process. The first band considers as input data and the second band is the target output. The predicted band is actual output data of BIP-MLPNN. Then the actual output band will be used as the input band for the next prediction process. That is to say, for the upcoming third band, the decoded second band (after compensating the residual with the predicted second band with help of decoded biases and weights) is used as the input. The actual third band is the training target. In the decoding site, we reconstruct the predicted band from the biases, weights and the previous band. Then the iterative process will continue until the last band has been trained and reconstructed at the decoder side.

*A. Motivation*

For demonstration, we take *cuprite 1* image from AVIRIS data set as an example. Fig. 3 shows that the average of the correlation coefficient (CC) [45] of the spectral inter-band can be up to 0.975 for the HS image.

A three-dimensional spatial Correlation Coefficient $P_{cc}$ can be evaluated using the equation (1),

$$P_{cc} = \frac{\sum_{k=1}^{N_k}\sum_{j=1}^{N_j}[P(k,j,l)-\overline{P_l}][P(k,j,l+1)-\overline{P_{l+1}}]}{\sqrt{\left(\sum_{k=1}^{N_k}\sum_{j=1}^{N_j}[P(k,j,l)-\overline{P_l}]^2\right)\left(\sum_{k=1}^{N_k}\sum_{j=1}^{N_j}[P(k,j,l+1)-\overline{P_{l+1}}]^2\right)}} \quad (1)$$

Where, one image data sample is *P (k,j,l)*,
$\{(k,j,l) | 1 \le k \le N_k, 1 \le j \le N_j, 1 \le l \le N_l\}$, $\overline{P_l}$ is the average pixels intensity of *l* spectral band.

A stronger correlation is assumed if the CC value is closer to 1. We can see a high degree of linear correlation in the figure. It provides the foundation for using the current band to approximate reflectance value of the succeeding band. A few damaged bands around band 101 and band 151 have been eliminated to assure HS data correctness.

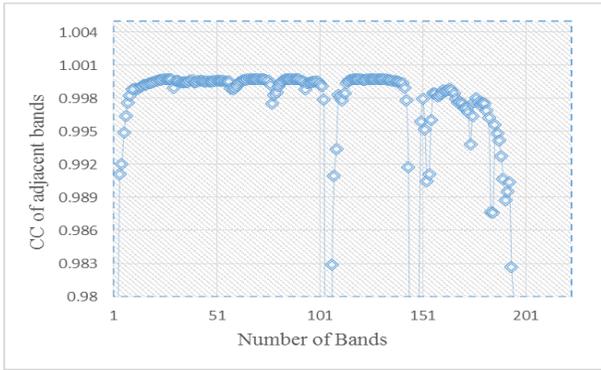

Fig. 3. The correlation coefficient of adjacent bands. The average of the correlation coefficient (CC) of the neighbouring band can be up to 0.975.

*Image Data Pre-process*

The data correlation among the neighbouring pixels e.g., between the current band *l* and the next *l+1* band is the basis for the proposed predicted coding technique. We will use BIP-MLPNN learning algorithm based feedforward neural network to predict *l+1* band, and after that encoding the residual. The goal is to find the minimized value of mean square error (MSE) of surrounding bands.

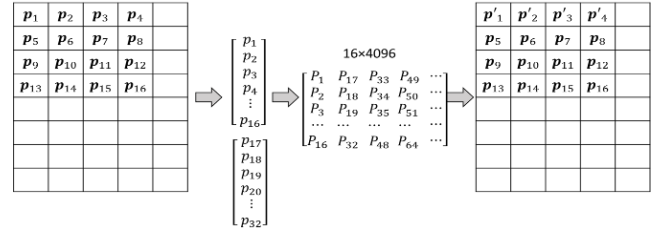

Fig. 4. Construction of input space. Input HS image are divided into 4×4 pixel blocks, and each block is converted to a 16×1 vector. The input data is *l* band and the output will be predicted *l+1* band.

Each band has been resized to a 256×256 matrix and rescaled to a range between [0, 1] that is MPLNN input and output expected values. Then the image is partitioned into 4×4 pixel blocks and each block is converted to a 16×1 vector, e.g., input neuron numbers are 16. Each band converts to a 16×4096 matrix as seen Fig. 4.

*Multilayer Neural Network Architecture*

The BIP-MLPNN is a 16-10-16 architecture layer. This architecture network can be used as a function approximator. The objective is to find a function that maps from the current band (input band) to approximate the reflectance value of the following band (target band). The input data will be a 16× 4096 matrix of *l* band and the output will be predicted (*l+1*) band. We employed a transfer function *tansig* as hidden layer and the linear function *purelin* as the output layer. Using *purelin* as output layer data there is no need for it to be normalised, however, normalised data can speed up the convergence rate. In the multilayer networks the output of one layer becomes the input to the following layer. The advantage of this structure of the network is that it can be used as a non-linear approximator and constrained the outputs of the network between 0 and 1. The network structure is illustrated in Fig. 5. The equations for the hidden layers $f^1(x)$ and output layer $f^2(x)$ is given at equation (2) where *x* is the net input to a neuron.

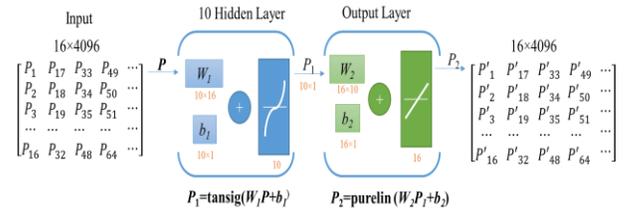

Fig. 5. MLPNN network structure is a 16-10-16 architecture layer. The tangent sigmoid transfer function is for the Hidden layers and linear function purelin is for the Output layers.

$$f^1(x) = \frac{2}{1+e^{-2x}} - 1 \text{ and } f^2(x) = x \quad (2)$$

*The Error Calculations and Weight Adjustments*

The training process of BIP-MLPNN is basically tuning the values of the weights and biases of the network to approximate the expected value. The training process first is to propagate the input forward through the network and then propagate the sensitives backward through the network. First, we need to choose some initial values for the weights and biases in the range from -1 and 1 before training the network. The initial values of the weight and the bias are chosen randomly. Using the above equation (2) to calculate input and output value of each layer respectively. The mean squared error $E$ between the target $P$ and the predictive value $P'$ is defined as:

$$E = \frac{1}{N_k N_j} \sum_{k=1}^{N_k} \sum_{j=1}^{N_j} (P_{(k,j,l+1)} - P'_{(k,j,l+1)})^2 \quad (3)$$

where $N_k \times N_j$ is the number of $P$. Then the weight vector $W$ can be updated by

$$W_{n+1} = W_n - [J^T J + \mu I]^{-1} J^T E \quad (4)$$

where $I$ is the identity matrix, $\mu$ giving as learning rate, $J$ is Jacobian Matrix [46]. Jacobian Matrix is easier to calculate, which doesn't need to calculate second-order partial derivatives. Therefore, we choose Levenberg-Marquardt (LM) algorithm as a training function. The LM generally is the fastest training function. It embraces the Gauss–Newton algorithm (GNA) and the method of gradient descent together.

*Bands reconstruction*

The only need to encode the weights $\{W_1\}$ $\{W_2\}$ and biases $\{b_1\}$ $\{b_2\}$ and send to BIP-MLPNN decoder side. BIP-MLPNN decoder is the same structure network as encoder side. We only use the previous $l$ band as an input band, substituted parameters $\{W_1\}$ $\{b_1\}$ and $\{W_2\{b_2\}$ into equation (2). Then we have:

$$\begin{cases} p_1 = f^1(x) = f^1(W_1 p + b_1) = \dfrac{2}{1 + e^{-2(W_1 p + b_1)}} \\ p_2 = f^2(x) = f^2(W_2 p_1 + b_2) = W_2 p_1 + b_2 \end{cases}$$

Where $p$ is input data, $p_1$ is the output from the hidden layers and output layer $p_2$ is the linear output layers. The target band $P_{l+1}$ can be calculated. For each 16×4096 band matrix, we got $n \times 16$ $\{W_1\}$ and $n \times 16$ $\{W_2\}$ matrix need to be encoded, n represents the number of hidden layers, e.g., each band with 65536 pixels, BIP-MLPNN model using10 hidden layers will only need to encode a $10 \times 16$ $\{W_1\}$, $16 \times 10$ $\{W_2\}$ matrix, and $10 \times 1$ $\{b_1\}$, $16 \times 1$ $\{b_2\}$ vector. We process matrices by mapping minimum and maximum values to [0, 255] data space then follow a lossless encoding algorithm. The compression ratio is therefore significantly improved. Let's see an example: a size of 256×256 image need to encode 65,536 pixels, if only the weights $\{W_1\}$ $\{W_2\}$and biases $\{b_1\}$ $\{b_2\}$, there will only encode two 16×10 matrixes for weights, 10×1 for the first bias and 16×1 for the second bias. Their values are from 0 to 255 after mapping. We also need to encode the maximum and minimum values of each matrix for mapping data. The experimental results reveal that the bit requirements for encoding the maximum, minimum, biases and weights are less than 1% data to be encoded, in comparison with encoding HS inter-band residuals.

*Prediction Mechanisms*

The predicted band result derived from the target values is further compensated in this step. The ultimate purpose of error-correction learning is bind the relative error between predicted band and the target band. The predicted band is approximated to the targeted band but is not identical, in some cases bias needs to be corrected. Because the decoder has no original pixel information, refer to Marco *et al.* [9], using quantization step sizes in a predictive lossy compression that could reach a near-lossless compression ratio. We also encode quantization step. The process is an adaptive prediction mechanism. Assume that $O_l$ is the compensation value for $l$ band, $O_l = [P_l(1 + \lambda)] - P_l^R$ if the predicted pixel value is smaller than the targeted band pixel. Or the predicted pixel is bigger, set $O_l = [P_l(1 - \lambda)] - P_l^R$. In the decoder, $O_l$ will be transmitted in the form of a short integer. This data will be read by the decoder. It is used to recalculate the image data after compensation $P'^R_l = P_l^R + O_l$ $q_l$ is the quantization step size applied to that residual. $P_l^R$ is the reconstructed pixel and $\lambda$ represents maximum acceptable relative reconstruction error.

DATASET

Airborne Visible/Infrared Imaging Spectrometer (AVIRIS) reflectance data is a publicly available high-dimensional HS datasets from NASA[1]. AVIRIS data are mainly collected for identifying, measuring, and monitoring constituents of the earth's surface and atmosphere based on molecular absorption and particle scattering signatures. It represents geological features of the earth. The 224 contiguous spectral bands have covered from 0.4 to 2.5 *μm* spectral range. The wavelength of each spectral band is approximately 10 *nm*. As seen in Fig. 6, *Cuprite 1, Cuprite 2, Cuprite 3 and Cuprite 4* were collected from mineral mapping at Cuprite, Nevada. Moffett Field is from California. Jasper Ridge is located in the central region of the Coast Range of California.

| Name | Imaging | Feature |
|---|---|---|
| Cuprite 1 | 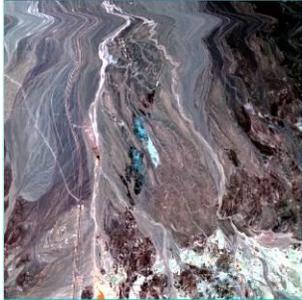 | RGB band # [76,143,180] Geological features, Mountain View |
| Cuprite 2 | 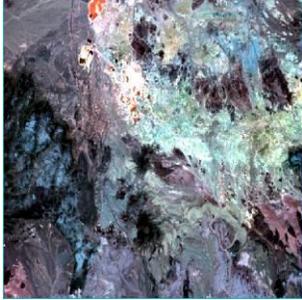 | RGB band # [76,143,180] Geological features |
| Cuprite 3 | 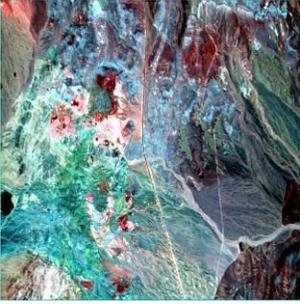 | RGB band # [43,143,170] Geological features with low variance |
| Cuprite 4 | 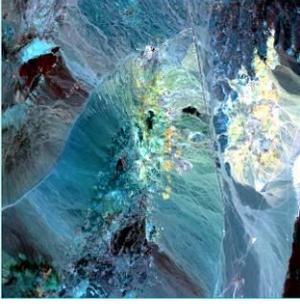 | RGB band # [43,143,180] geological features with high variance |
| Moffett | 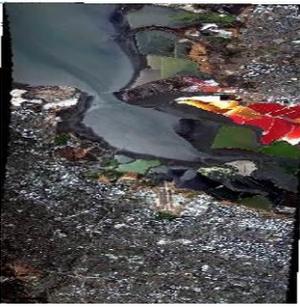 | RGB band # [76,143,180] vegetation, urban, water |
| Jasper | 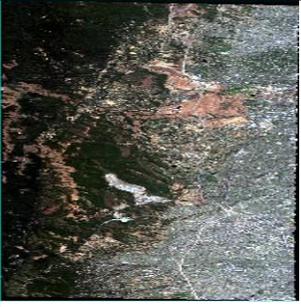 | RGB band # [29,20,12] Vegetation distributions wetlands, grasslands and serpentine soils |

Fig. 6. AVIRIS reflectance imaging demonstration.

EXPERIMENT RESULT

All of the experimental AVIRIS images have been resized to 256×256 for verification purpose. The numbers of neurons of input and output layers are 16. The maximum time of training in second is 1000. The training goal of MSE is set to 0.0001 The validation of BIP-MLPNN model is obtained at the end of each training procedure. MSE gives the difference between observation and simulation. The performance of training and testing plots the progress in Fig. 7 which indicated that the validation and test curves are very similar. The iteration at 219 Epochs performance reached the minimum MSE. Fig. 8 creates four regression plots for the training set (Train), the validation set (Validation), the test set (Test) and the entire dataset (All).

It illustrates the obtained output, target and R-values. The correlation coefficient (R-value) between the outputs and the targets values is well fitted. R-values calculated for evaluating the trained BIP-MLPNN model. The reflectance values of pixels are plotted against the targets (circles). The dashed line in each plot represents the perfect result – outputs = targets. The solid line indicates the best linear fit. The R-value related to the training set is very close to 1, which indicates a very good fit. The R-value related to the test set is 0.99636, the training and validation results also show R values that are greater than 0.99.

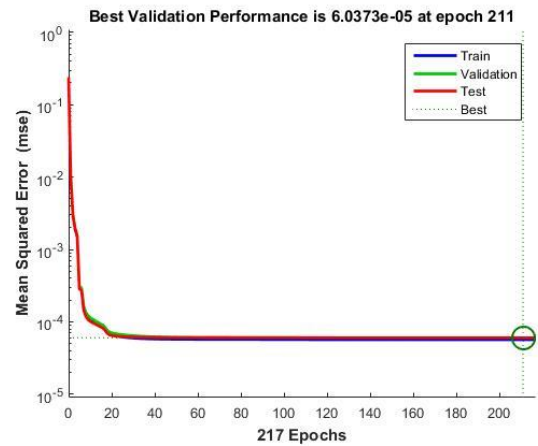

Fig. 7. The validation performance of the BIP-MLPNN model.

Fig. 9 is a demonstration of the compress results using band 101 ($l$=100) of the HS image cuprite 1. The predicted $l$+1 band is after prediction of the proposed method and the residual between the $l$ and $l$+1 neighbouring bands. The residual images calculated the prediction error between target and predicted $l$+1 band are shown in its histogram. The figure is illustrated that most of the residual values are 0 which indicates the BIP-MLPNN approximator achieves good results, therefore, result in requiring less number of bits to encode.

Fig. 10 shows rate distortion (RD) performance of selected experimental HS images using the proposed BIP-MLPNN model, JP2K, JPEG2K-residual, JPEG, JPEG-residual encoder techniques. The reason we select those four encoders for comparison, mainly because they all are very well-known reliable benchmark encoders. At present, many new developed methods are actually still based on these algorithms with partial improvements. We may present more accurate and reliable results by comparison with those benchmarking encoders. JP2K is treating each band individually, encoding each band separately by using JPEG 2000 (JP2k) image compression standard and coding system. JPEG is the similar method to deal with each band but using the JPEG standard. JP2K-residual and JPEG-residual are to calculate residual of current band and the previous band, then only encode residual instead of encoding each band by JP2K or JPEG encoder. Besides, we also choose other competitive encoders, the state-of-the-art methods: 3DSPECK [27] AT-3DSPECK [28], AT-3DSPIHT [20] for comparison purpose in Fig. 10.

BIP-MLPNN model shows obvious advantages at low bit rate ranges as it only needs to encode weights and biases. Both JPEG2K-residual and JPEG-residual are encoded based on the residuals between the target band $l+1$ and the previous band $l$. Overall, BIP-MLPNN model improved the compression ratio at very low bit rate.

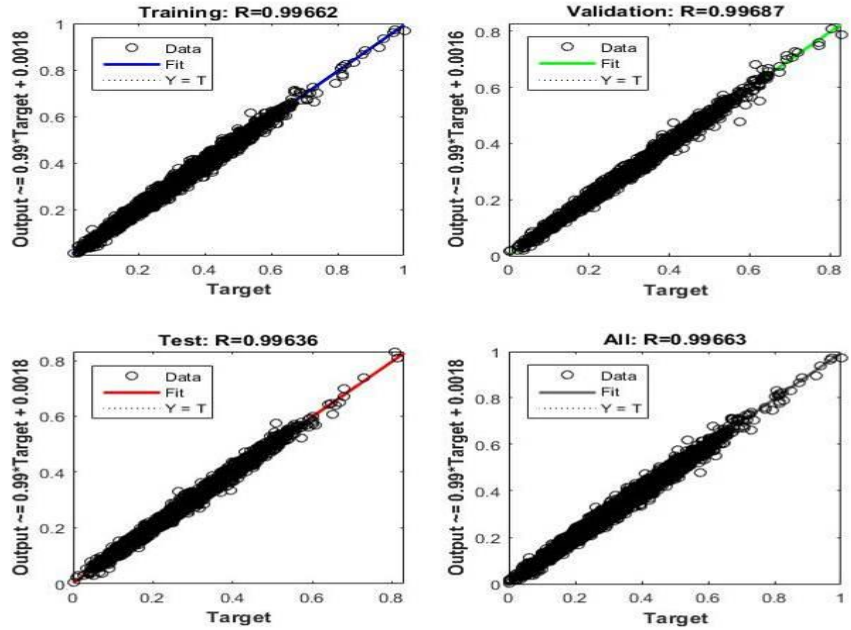

Fig. 8. Performance of the BIP-MLPNN model, four regression plots for the training set (Train), the validation set (Validation), the test set (Test) and the entire dataset.

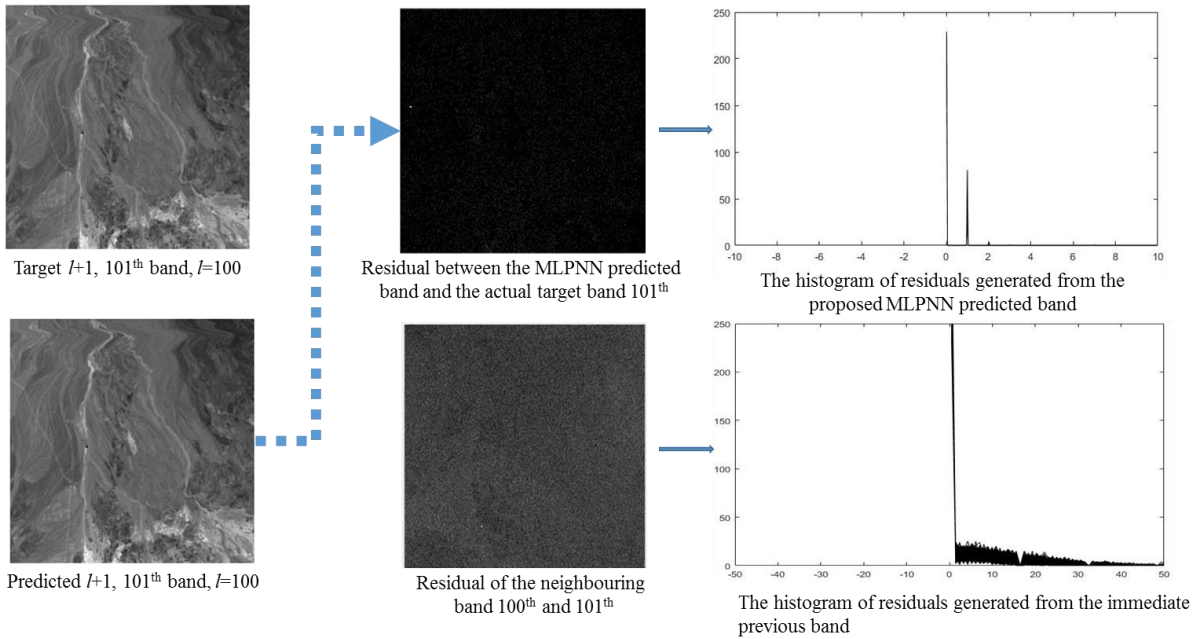

Fig. 9. A demonstration of HS image cuprite 1 compressed result by MLPNN model.

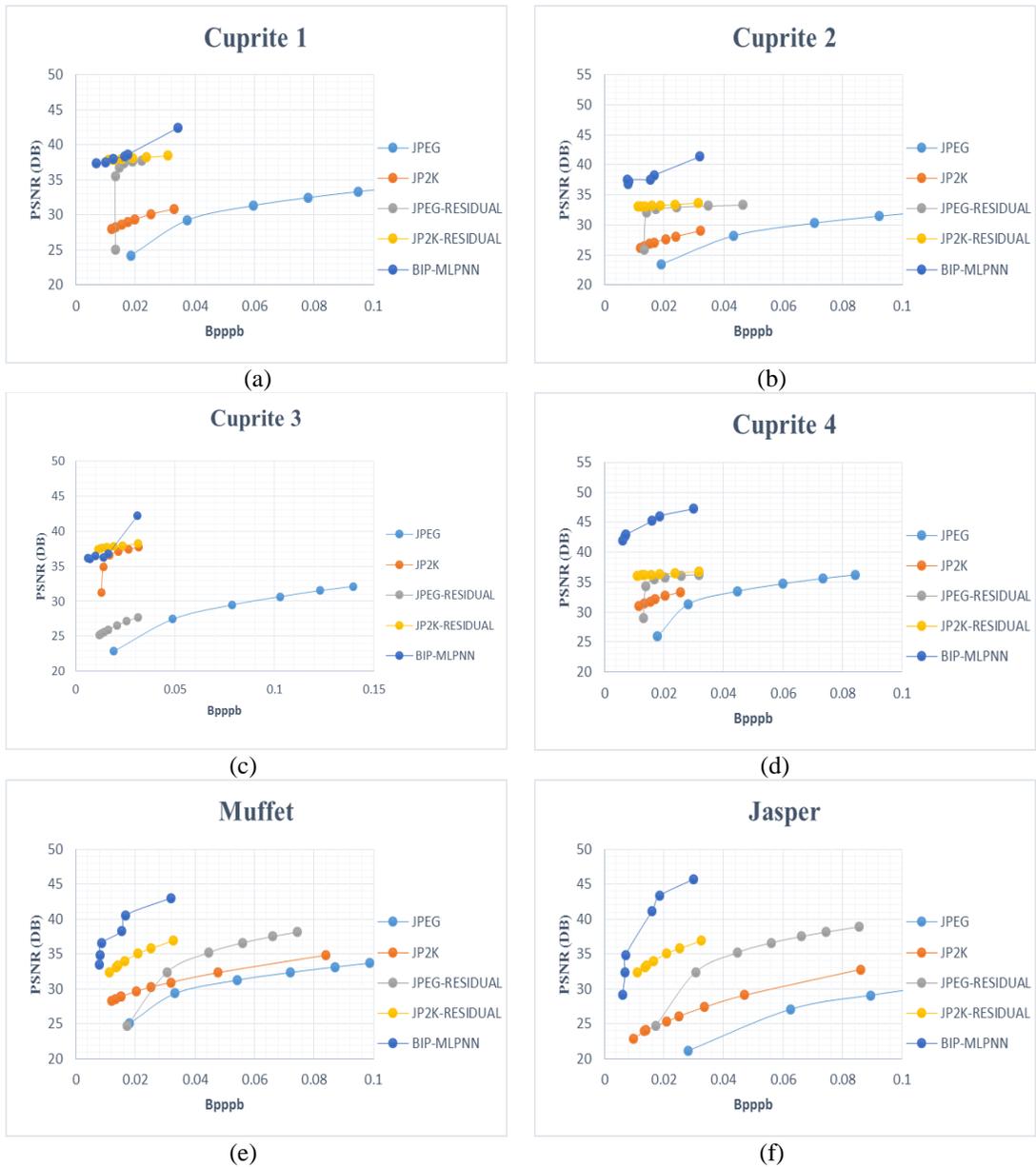

Fig. 10. Rate distortion performance of HS images using the proposed BIP-MLPNN model, JP2K (JPEG 2000), JP2K-residual (JPEG 2000-residual), JPEG, JPEG-residual encoder techniques and further comparison at low bit rates

TABLE 1 PREDICTOR PERFORMANCE COMPARISON WITH CCSDS-123 STANDARD

|  | MSE | | SSIM | | PSNR | |
| --- | --- | --- | --- | --- | --- | --- |
|  | **BIP-MLPNN** | **CCSDS 123** | **BIP-MLPNN** | **CCSDS123** | **BIP-MLPNN** | **CCSDS 123** |
| Cuprite 1 | **0.2562** | 0.5676 | **0.9997** | 0.9986 | **54.0456** | 50.5903 |
| Cuprite 2 | **0.6857** | 0.7736 | **0.999** | 0.9996 | **49.7692** | 49.2454 |
| Cuprite 3 | 0.627 | **0.6189** | 0.9988 | **0.9996** | 50.158 | **50.2145** |
| Cuprite 4 | **0.3074** | 0.4634 | **0.9987** | 0.9984 | **53.2537** | 51.4715 |
| Muffet | **0.1892** | 0.6702 | **0.9996** | 0.9991 | **55.3607** | 49.8686 |
| Jasper | **0.0120** | 0.0185 | **0.9972** | 0.9959 | **43.7339** | 41.8769 |

To further clarify the accuracy of the BIP-MLPNN predictor, the experiments use three metrics the MSE, the Structural Similarity index (SSIM) and the Peak signal-to-noise ratio (PSNR) for a comparative analysis with the CCSDS-123 predictor. This experiment is focusing on the accuracy of the prediction, no entropy coding involved. The MSE, the SSIM and the PSNR are calculated between the target band and the predicted band in Table 1. Through the comparison of experimental results，BIP-MLPNN predicted data can fit target data well. It shows more advantages on both the MSE and PSNR metrics. On the other hand, both predictors get very close values at SSIM metrics, all above 0.998. CCSDS-123 predicted data get slightly higher value than the BIP-MLPNN in *Cuprite 3*. Analysis on the differences of HS data, we noticed that the image is computed from three different terms: luminance (mean), contrast (variance) and structure (correlation). The metric can be insensitive probably because large regions of low variance exists. After comparing the variance of *Cuprite 1* and *Cuprite 3*. We may see that the *Cuprite 3* variance is about 109.7543, but variance of *Cuprite 1* is up to 217.

To compare with the state-of-the-art methods 3DSPECK [27] AT-3DSPECK [28], AT-3DSPIHT [20], GRNN [43], we show the image quality against three different bpppbs at very low bit rate e.g., 0.2, low bit rate 0.5 and high rate 1.0. The result shown in Fig. 11 has testified that the BIP-MLPNN has certain advantages at very low bit rate range i.e. less than 0.2 bpppb. This indicated that the BIP-MLPNN can achieve a relative higher compression ratio with very low bit rate.

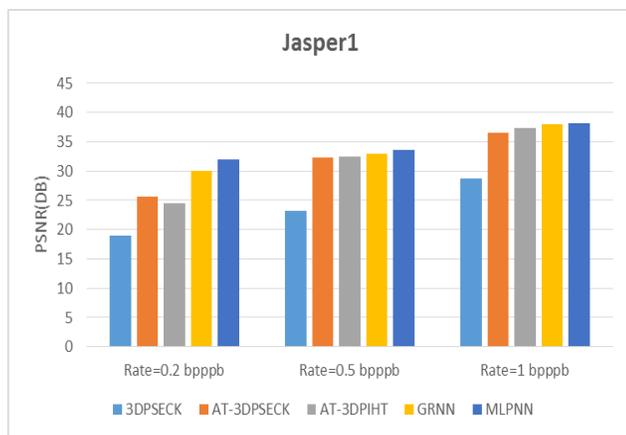

Fig. 11. Performance comparison for Jasper ridge scene at rate 0.2, 0.5 and 1 bpppb

When 3D set partitioning coding methods in hyperspectral image compression performance reached the minimum error, the training time is also calculated which is shown in Fig. 12. Each training session per spectral band, the longest one is Moffett which costs about 30 second/band and the shortest training time recorded for Cuprite 04 is 15 second/band. Note that the experimental is performed in a PC with Intel® Core™ i5-4760 and RAM 8GB.

## CONCLUSION

A new lossy block-based predictive BIP-MLPNN model which makes use of the multilayer neural network to predict the succeeding bands has been proposed. In the model, the current spectral band is used as a training data set, and the next band as the training target set on the encoder side. Each band is converted to a 16×4096 matrix. The experimental results show that the proposed technique outperforms the JPEG and JPEG2K as well as 3DSPECK, AT-3DSPECK, AT-3DSPIHT, according to the rate distortion performance at very low bit rates where the bpppb range is below 0.2.

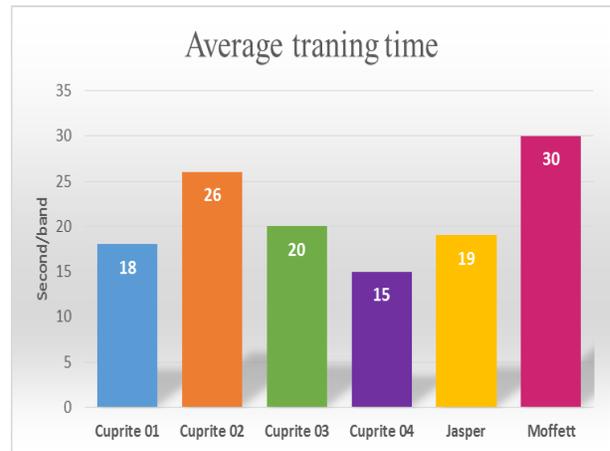

Fig. 12. Average training time second/band

The major advantage of this approach is that it only needs to use the binary entropy encoder to encode weights and biases; later in the decoder side, the target band can be reconstructed by using transferred weights, biases, and residual. Therefore, only a few bits of data need to be transferred. The obtained results were found to be quite satisfactory. The next challenge would be an implementation of deep learning neural networks to further improve the accuracy of the predictive model.

REFERENCES

[1] "NASA's Airborne Visible/Infrared Imaging Spectrometer," https://aviris.jpl.nasa.gov/data/free_data.html.
[2] G. Cheng, and J. Han, "A survey on object detection in optical remote sensing images," *ISPRS Journal of Photogrammetry and Remote Sensing,* vol. 117, pp. 11-28, 2016.
[3] I. Makki, R. Younes, C. Francis *et al.*, "A survey of landmine detection using hyperspectral imaging," *ISPRS Journal of Photogrammetry and Remote Sensing,* vol. 124, pp. 40-53, 2017.
[4] S. Munera, C. Besada, N. Aleixos *et al.*, "Non-destructive assessment of the internal quality of intact persimmon using colour and VIS/NIR hyperspectral imaging," *LWT-Food Science and Technology,* vol. 77, pp. 241-248, 2017.
[5] R. Kemker, and C. Kanan, "Self-taught feature learning for hyperspectral image classification," *IEEE Transactions on Geoscience and Remote Sensing,* vol. 55, no. 5, pp. 2693-2705, 2017.
[6] J.-H. Cheng, B. Nicolai, and D.-W. Sun, "Hyperspectral imaging with multivariate analysis for technological parameters prediction and classification of muscle foods: A review," *Meat science,* vol. 123, pp. 182-191, 2017.
[7] B. Park, and R. Lu, *Hyperspectral Imaging Technology in Food and Agriculture*: Springer New York, 2015.
[8] C.-I. Chang, *Hyperspectral data processing: algorithm design and analysis*: John Wiley & Sons, 2013.

[9] D. Zhao, S. Zhu, and F. Wang, "Lossy hyperspectral image compression based on intra-band prediction and inter-band fractal encoding," *Computers & Electrical Engineering,* vol. 54, pp. 494-505, 2016.

[10] S. Zhu, D. Zhao, and F. Wang, "Hybrid prediction and fractal hyperspectral image compression," *Mathematical Problems in Engineering,* vol. 2015, 2015.

[11] H. Shen, W. D. Pan, and D. Wu, "Predictive Lossless Compression of Regions of Interest in Hyperspectral Images With No-Data Regions," *IEEE Transactions on Geoscience and Remote Sensing,* vol. 55, no. 1, pp. 173-182, 2017.

[12] F. Rizzo, B. Carpentieri, G. Motta et al., "Low-complexity lossless compression of hyperspectral imagery via linear prediction," *IEEE Signal Processing Letters,* vol. 12, no. 2, pp. 138-141, 2005.

[13] J. Mielikainen, and B. Huang, "Lossless compression of hyperspectral images using clustered linear prediction with adaptive prediction length," *IEEE geoscience and remote sensing letters,* vol. 9, no. 6, pp. 1118-1121, 2012.

[14] G. Motta, F. Rizzo, and J. A. Storer, *Hyperspectral data compression*: Springer Science & Business Media, 2006.

[15] A. C. Bovik, *Handbook of image and video processing*: Academic press, 2010.

[16] B. Aiazzi, S. Baronti, and L. Alparone, "Lossless compression of hyperspectral images using multiband lookup tables," *IEEE Signal Processing Letters,* vol. 16, no. 6, pp. 481-484, 2009.

[17] J. Mielikainen, and P. Toivanen, "Lossless compression of hyperspectral images using a quantized index to lookup tables," *IEEE Geoscience and Remote Sensing Letters,* vol. 5, no. 3, pp. 474-478, 2008.

[18] G. G. King, C. C. Seldev, and N. A. Singh, "A Novel Compression Technique for Compound Images Using Parallel Lempel-Ziv-Welch Algorithm," *Applied Mechanics and Materials,* vol. 626, pp. 44, 2014.

[19] P. G. Howard, and J. S. Vitter, "New methods for lossless image compression using arithmetic coding," *Information processing & management,* vol. 28, no. 6, pp. 765-779, 1992.

[20] L. Sasilal, and V. Govindan, "Arithmetic Coding-A Reliable Implementation," *International Journal of Computer Applications,* vol. 73, no. 7, 2013.

[21] N. Amrani, J. Serra-Sagristà, V. Laparra et al., "Regression wavelet analysis for lossless coding of remote-sensing data," *IEEE Transactions on Geoscience and Remote Sensing,* vol. 54, no. 9, pp. 5616-5627, 2016.

[22] S. Shahriyar, M. Paul, M. Murshed et al., "Lossless Hyperspectral Image Compression Using Binary Tree Based Decomposition." pp. 1-8.

[23] M. Paul, R. Xiao, J. Gao et al., "Reflectance Prediction Modelling for Residual-Based Hyperspectral Image Coding," *PloS one,* vol. 11, no. 10, pp. e0161212, 2016.

[24] X. Tang, and W. A. Pearlman, "Three-dimensional wavelet-based compression of hyperspectral images," *Hyperspectral Data Compression*, pp. 273-308: Springer, 2006.

[25] M. Zala, and S. Parmar, "3D Wavelet transform with SPIHT algorithm for image compression," *International Journal of Application Or Innovation in Engineering & Management (IJAIEM),* vol. 2, no. 5, 2013.

[26] H. H. Zayed, S. E. Kishk, and H. M. Ahmed, "3D wavelets with SPIHT coding for integral imaging compression," *Int J Comput Sci Netw Secur,* vol. 12, no. 1, pp. 126, 2012.

[27] X. Tang, S. Cho, and W. A. Pearlman, "3D set partitioning coding methods in hyperspectral image compression." pp. II-239.

[28] X. Wu, and N. Memon, "Context-based lossless interband compression-extending CALIC," *IEEE Transactions on Image Processing,* vol. 9, no. 6, pp. 994-1001, 2000.

[29] G. Ulacha, and R. Stasiński, "New context-based adaptive linear prediction algorithm for lossless image coding." pp. 1-4.

[30] E. Magli, G. Olmo, and E. Quacchio, "Optimized onboard lossless and near-lossless compression of hyperspectral data using CALIC," *IEEE Geoscience and remote sensing letters,* vol. 1, no. 1, pp. 21-25, 2004.

[31] F. L. Hitchcock, "The expression of a tensor or a polyadic as a sum of products," *Studies in Applied Mathematics,* vol. 6, no. 1-4, pp. 164-189, 1927.

[32] L. R. Tucker, "Some mathematical notes on three-mode factor analysis," *Psychometrika,* vol. 31, no. 3, pp. 279-311, 1966.

[33] A. Karami, R. Heylen, and P. Scheunders, "Hyperspectral Image Compression Optimized for Spectral Unmixing," *IEEE Transactions on Geoscience and Remote Sensing,* vol. 54, no. 10, pp. 5884-5894, 2016.

[34] R. A. Harshman, "Foundations of the PARAFAC procedure: Models and conditions for an" explanatory" multi-modal factor analysis," 1970.

[35] J. D. Carroll, and J.-J. Chang, "Analysis of individual differences in multidimensional scaling via an N-way generalization of "Eckart-Young" decomposition," *Psychometrika,* vol. 35, no. 3, pp. 283-319, 1970.

[36] L. Zhang, L. Zhang, D. Tao et al., "Compression of hyperspectral remote sensing images by tensor approach," *Neurocomputing,* vol. 147, pp. 358-363, 2015.

[37] M. A. Veganzones, J. E. Cohen, R. C. Farias et al., "Nonnegative tensor CP decomposition of hyperspectral data," *IEEE Transactions on Geoscience and Remote Sensing,* vol. 54, no. 5, pp. 2577-2588, 2016.

[38] B. Du, M. Zhang, L. Zhang et al., "PLTD: Patch-based low-rank tensor decomposition for hyperspectral images," *IEEE Transactions on Multimedia,* vol. 19, no. 1, pp. 67-79, 2017.

[39] B. Du, Z. Huang, and N. Wang, "A Bandwise Noise Model Combined With Low-Rank Matrix Factorization for Hyperspectral Image Denoising," *IEEE Journal of Selected Topics in Applied Earth Observations and Remote Sensing,* vol. 11, no. 4, pp. 1070-1081, 2018.

[40] B. Du, Z. Huang, N. Wang et al., "Joint weighted nuclear norm and total variation regularization for hyperspectral image denoising," *International Journal of Remote Sensing,* vol. 39, no. 2, pp. 334-355, 2018.

[41] H. Faris, I. Aljarah, and S. Mirjalili, "Training feedforward neural networks using multi-verse optimizer for binary classification problems," *Applied Intelligence,* vol. 45, no. 2, pp. 322-332, 2016.

[42] O. N. A. AL-Allaf, "Fast Backpropagation Neural Network algorithm for reducing convergence time of BPNN image compression." pp. 1-6.

[43] K. K. Halder, and M. Paul, "Interband prediction of hyperspectral images using generalized regression neural network." pp. 175-179.

[44] M. T. Hagan, H. B. Demuth, and M. H. Beale, "Neural network design, PWS Pub," *Co., Boston,* vol. 3632, 1996.

[45] C.-h. Chen, *Signal and image processing for remote sensing*: CRC press, 2012.

[46] I. S. Gradshteyn, and I. M. Ryzhik, *Table of integrals, series, and products*: Academic press, 2014.

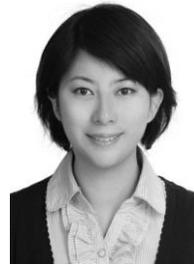

Rui Dusselaar received the B.S. degree in Computer Information System Management from Xi'an International University, China, in 2001 and the M.S. degree in Electronic Engineering from University of Malaysia Pahang, in 2013. She is currently pursuing the PhD degree at Charles Sturt University, Australia. Her major research interests are in the field of Multidimensional data compression, such as Hyperspectral image compression, Image processing, Machine Learning.

From 2001 until 2010, she has worked as a programmer for the IBM Tivoli Netcool/Reporter 2.2.0.0/2.1 system. Mrs Dusselaar has mostly participated in the Business intelligence software Viador BI 7.0, development, Beijing forum Project and electronic multi-tier clinical information system. Then she awards scholarship, the UMP Graduate Research Scheme and Postgraduate research grant scheme for her master study. Currently, she is receiving Higher Degree by Research Scholarship from Charles Sturt University.

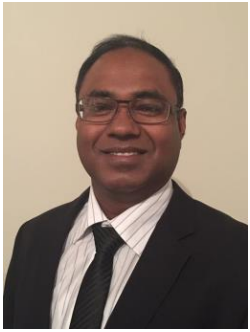 Manoranjan Paul received PhD degree from Monash University, Australia in 2005. He was a Post-Doctoral Research Fellow in the University of New South Wales, Monash University, and Nanyang Technological University. Currently he is an Associate Professor and Director of E-Health Research at Charles Sturt University (CSU).

His major research interests are in the field of Data Science such as Data Compression, Video Technology, Computer Vision, E-Health, and Medical Imaging/Signal Processing. He has published more than 150 refereed publications. He was an invited keynote speaker in DICTA 2017 & 2013, CWCN 2017, WoWMoM 2014, and ICCIT 2010.

Currently, he is as an Associate Editor of IEEE Transactions on Circuits and Systems for Video Technology and EURASIP Journal in Advances on Signal Processing. He was a Program Chair of PSIVT 2017 and Publicity Chair of DICTA 2016.

He is the ICT Researcher of the Year 2017 awarded by Australian Computer Society. He obtained Research Excellence Supervision and Research Excellence Awards at CSU. He obtained more than $15M competitive grant money including the most prestigious Australian Research Council (ARC) Discovery Project Grant, Cybersecurity CRC.